# Vehicle-Mounted Mid-Infrared Dual-Comb Spectroscopy for On-Road Trace Gas Detection

Xutian Jing†, Kaiwen Wei†, Chenglin Gu*, Xiong Qin, Junwei Li, Xingyin Yang, Zhaoting Huang, Jianping Zhang, Chenhao Sun, Chenyu Liu, Zejiang Deng, Zhiwei Zhu, Daping Luo*, Wenxue Li*, and Heping Zeng

†Xutian Jing and Kaiwen Wei contributed equally to this work.

*Correspondence: clgu@lps.ecnu.edu.cn (C.G.); dpluo@lps.ecnu.edu.cn (D.L.); wxli@phy.ecnu.edu.cn (W.L.)

State Key Laboratory of Precision Spectroscopy, East China Normal University, Shanghai 200062, China

## Abstract

Advances in mid-infrared (MIR) dual-comb spectroscopy (DCS) have significantly enhanced molecular detection in recent years. The capability of DCS to precisely identify and quantify atmospheric trace gases makes it attractive for field applications across the environmental, agricultural, energy, and industrial sectors. In particular, there is a growing demand for mobile and continuous gas monitoring in outdoor environments where emission sources and sinks are often episodic and spatially heterogeneous. However, the practical field-deployment of DCS on mobile platforms under realistic field conditions has remained limited. This study demonstrates for the first time a vehicle-mounted MIR DCS system that enables continuous mobile atmospheric sampling across multiple outdoor sites and road environments. The system exhibited a stable signal-to-noise performance during on-road operation, including expressway driving at speeds up to 100 km/h. Furthermore, natural-gas leakage sources were successfully located and a two-dimensional methane concentration field was reconstructed around a controlled release source. In the future, the system can be integrated into more mobile platforms, such as unmanned aerial vehicles, enabling flexible trace gas detection over urban-scale regions.

## Introduction

As a tool for precise time-frequency measurement, the advantages of optical frequency comb technology have been demonstrated in many areas since its inception [1–4]. DCS combines broadband coverage, high-frequency accuracy, and rapid acquisition by mapping optical frequencies into the radio-frequency domain using two combs with slightly detuned repetition rates [5,6]. Operating in the MIR region, where most light molecules exhibit intense fundamental rotovibrational band absorption, DCS provides strong sensitivity and high precision even with relatively short optical paths [7,8]. Conventional outdoor gas-sensing techniques are well established, such as Fourier transform spectrometry and tunable diode laser absorption spectroscopy; however, they often involve trade-offs between spectral coverage, temporal resolution, sensing distance, and dependence on solar illumination, particularly in complex atmospheric environments [9–11]. These characteristics make MIR DCS highly promising for outdoor gas sensing across environmental, agricultural, and energy-related applications [12–16].

In recent years, pioneering efforts, particularly those led by the National Institute of Standards and Technology and their collaborators, have explored the extension of DCS from controlled laboratory setups to outdoor open-path measurements. Early demonstrations achieved the first outdoor near-infrared DCS over kilometer-scale open paths, establishing DCS as a precise approach for continuous monitoring of regional greenhouse gas concentrations under ambient atmospheric conditions [17]. Subsequent work extended DCS to spatially-resolved atmospheric measurements using a multicopter-mounted retroreflector [18]. Further advances introduced the first field-deployed DCS systems capable of continuously detecting and quantifying trace-gas emission sources over multi-square-kilometer regions, bridging continuous but localized monitoring and quasi-instantaneous "snapshot-in-time" regional monitoring [19]. Building on these efforts, open-path DCS was extended into the MIR for the first time, enabling kilometer-scale field measurements of volatile organic compounds (VOCs) [20]. Subsequently, the MIR DCS was deployed in the field to measure ambient VOCs along a 2 km open path [21]. In addition, Westberg et al. [22] from Princeton University demonstrated urban open-air chemical sensing using a quantum cascade laser dual-comb spectrometer, enabling field-deployed measurements of multiple trace gas species over retroreflector-based open paths in urban environments.

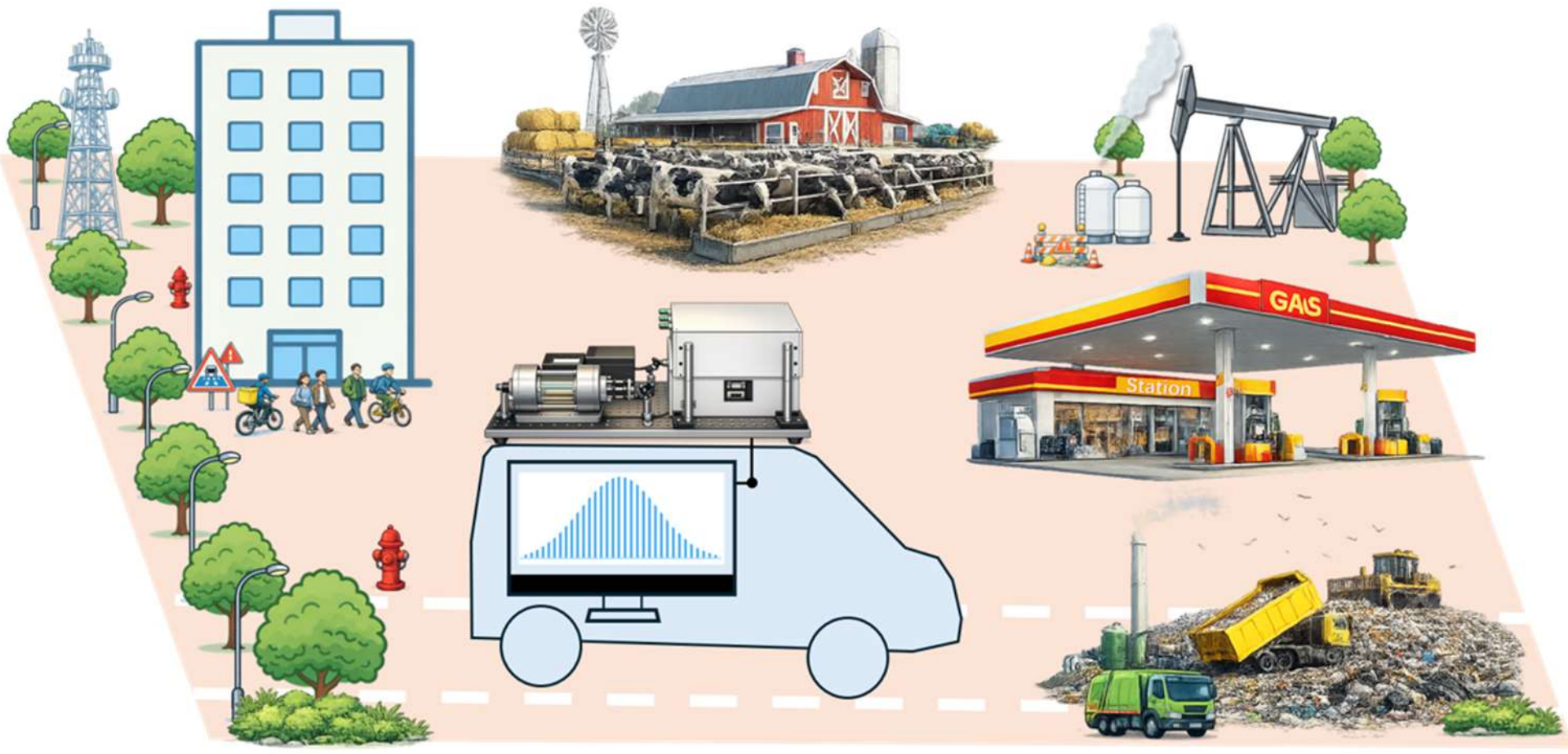

**Fig. 1**. Conceptual illustration of mobile methane monitoring in urban environments using the vehicle-mounted platform. The system enables on-road detection of methane emissions from representative sources, such as livestock farms, landfills, natural-gas stations, and oil and gas facilities.

Collectively, these works have demonstrated that field-deployable DCS has progressively evolved to deliver increasingly broad practical utility in outdoor sensing. Despite these impressive demonstrations, most field-deployed DCS systems to date rely on fixed telescope-retroreflector configurations, requiring preinstalled optical paths and yielding path-averaged concentrations along static lines of sight. Many emerging urban and industrial monitoring scenarios involve emission sources that are spatially heterogeneous, dynamically evolving, or not known a priori, motivating the development of mobile DCS platforms with enhanced spatial flexibility. However, implementing DCS on a mobile platform imposes substantially more stringent requirements on system-level miniaturization and portability. In addition, DCS relies on maintaining sub-wavelength-level phase stability and mutual coherence between the two combs. As a result, translating DCS from laboratory or fixed outdoor installations into a fully mobile platform remains technically challenging.

This study demonstrates a vehicle-mounted patrol-type MIR DCS system operating spanning 3.35–3.47 μm for the concentration detection of methane ($CH_4$) and water ($H_2O$) in urban environments, as illustrated in Fig. 1. Two key design strategies were employed to address the challenges associated with mobile DCS operations under outdoor environmental perturbations. First, all polarization-maintaining (PM) figure-9 oscillators were used to provide enhanced mechanical robustness for mobile deployment. Second, the optical-optical modulated frequency combs technique was employed to passively maintain mutual coherence between the two combs. The resulting system achieved a figure of merit of $3.4 \times 10^6 \sqrt{\text{Hz}}$, which is comparable to the performance of typical laboratory-based MIR DCS systems [6,23,24]. The system operated at an acquisition rate of 400 Hz, enabling a high temporal resolution under mobile conditions. To verify the capability of the system, patrol measurements of atmospheric $CH_4$ and $H_2O$ concentrations were conducted on the East China Normal University (ECNU) campus. In addition, a long-distance urban patrol was performed for approximately 1 h to evaluate the continuous operation under realistic driving conditions. The vehicle traveled on urban roads and highways at a maximum speed of 100 km/h. Furthermore, controlled natural-gas release experiments were conducted. In a subsequent experiment, two-dimensional concentration field mapping was performed around the release sites, and the measured plume distributions showed a strong correlation with wind-rose analyses. In summary, these results indicate that the vehicle-mounted patrol-type DCS technology can be used for trace gas monitoring, plume detection, and leakage source localization in actual field scenarios.

## Results and discussion

### Mid-infrared dual-comb spectroscopy

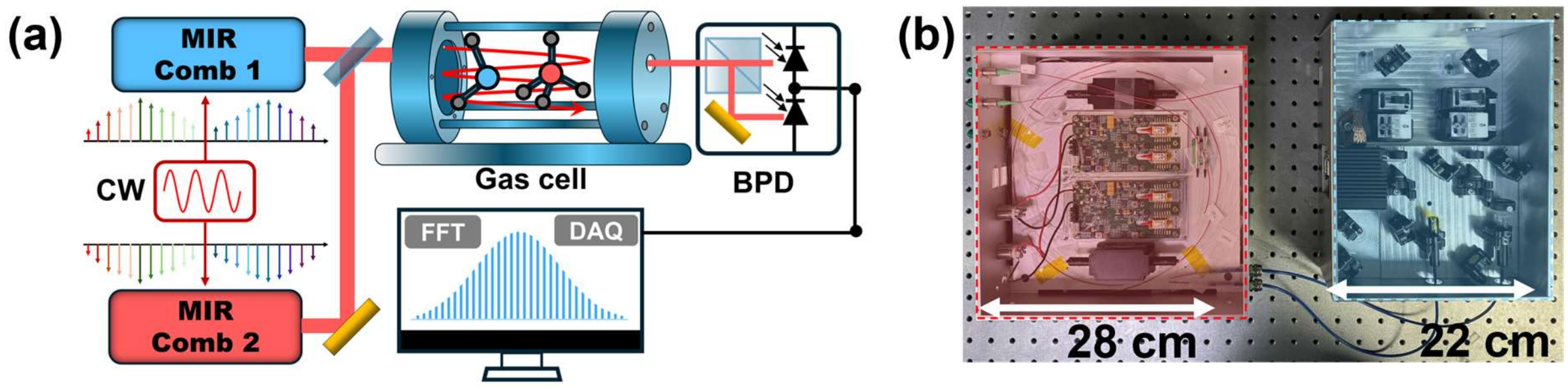

**Fig. 2**. Overview of the vehicle-mounted MIR DCS system. (a) Schematic of the integrated optical-optical modulated dual-comb spectrometer. The combined combs propagated through an open gas cell and were detected by a balanced photodetector (BPD). CW, continuous-wave; FFT, fast Fourier transform; DAQ, data acquisition. (b) Photograph of the integrated dual-comb system modules, highlighting the fiber-laser and frequency-conversion units. The red-shaded region shows the two fiber mode-locked lasers together with their amplifiers, while the blue-shaded region corresponds to the frequency-conversion module.

As illustrated in Fig. 2(b), the dual-comb spectrometer consisted of two main subsystems: (1) two mode-locked lasers and their amplifiers, and (2) two nonlinear frequency-conversion modules based on CW (continuous-wave)-seeded optical parametric generation (OPG) [25]. Two home-built ytterbium (Yb)-doped all PM fiber mode-locked lasers based on the nonlinear amplifying loop mirror technique generated ultrafast pulse trains centered at 1030 nm, operating at a repetition rate $f_r$ of 56.5 MHz with a repetition-rate difference $\Delta f_r$ of approximately 400 Hz. The all PM figure-9 cavity configuration has been reported to provide excellent low-noise performance and robust mechanical stability against environmental vibrations [26–29], making it suitable for operation under field conditions. The two free-running mode-locked lasers were co-packaged in an integrated enclosure and therefore experienced similar temperature fluctuations and mechanical vibrations. Consequently, the repetition rates of the two frequency combs experienced similar jitter, leading to reduced fluctuations in their repetition rate difference. The pulse trains were subsequently amplified using single-stage fiber amplifiers, delivering a maximum average output power of 550 mW with a pulse duration of approximately 4 ps at a central wavelength of 1030 nm. The amplified pulses served as the pump source for the subsequent CW-seeded OPG process.

A mutually coherent dual-comb was generated by directly pulsing the CW laser using an optical modulation technique, which was realized via the CW-seeded OPG process, as shown in Fig. 2(a). A CW laser centered at 3350 nm was split into two branches and used as the seed for the two CW-seeded OPG processes. In each branch, the CW seed laser was combined with the amplified pulses from the corresponding Yb-doped fiber laser and coupled into a MgO-doped periodically poled lithium niobate waveguide, where CW-seeded OPG occurred, generating MIR dual-comb spectra spanning 3.35–3.47 μm. Both MIR combs were referenced to a common CW seed laser, resulting in excellent passive mutual coherence [30,31]. At a pump power of 500 mW, each MIR comb branch provided an average output power of approximately 4 mW. The power can be further increased by compressing the pump pulses, which enhances the peak power and thereby improves the efficiency of the nonlinear frequency-conversion process [25]. The two MIR combs were then combined and coupled into an open Herriott gas cell with an effective optical path length of 25 m, thereby enhancing the detection sensitivity through extended light-molecule interactions. The resulting multi-heterodyne signal was detected by a balanced photodetector (Vigo, PVI-4TE-5), yielding one interferogram (IGM) every 2.5 ms. The IGMs were subsequently digitized using a 12-bit high-speed data acquisition card (AlazarTech, ATS9350) for FFT-based spectral retrieval.

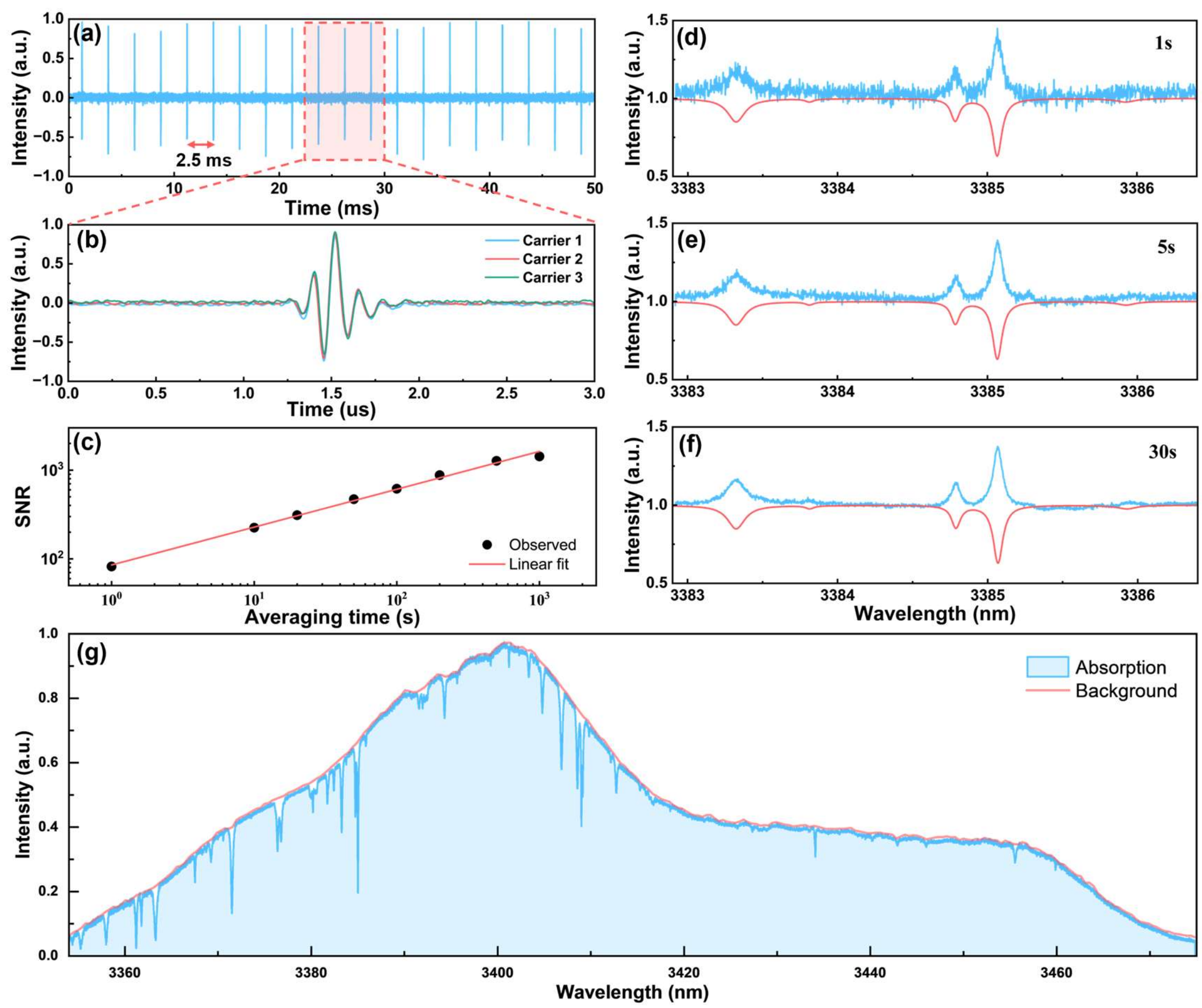

**Fig. 3.** Dual-comb spectroscopy of $CH_4$ and $H_2O$. (a) Time-domain multi-heterodyne signal generated by two MIR frequency combs, showing periodic IGMs with a recurrence period of $1/\Delta f_r$=2.5 ms. (b) Expanded views of three representative center bursts of the dual-comb IGMs within the same time window. (c) Spectral average signal-to-noise ratio (SNR) as a function of the averaging time, exhibiting a power-law scaling with an exponent of approximately 0.43. (d–f) Comparison of measured $H_2O$ absorption spectra (blue) with the HITRAN 2020 reference spectra (red) at averaging times of 1 s, 5 s, and 30 s, respectively. (g) Measured broadband absorption spectrum over the full spectral range with an averaging time of 10 s.

Fig. 3 summarizes the time-domain IGMs and the retrieved absorption spectra obtained by the MIR dual-comb system. Fig. 3(a) shows 20 IGMs recorded within a 50 ms time window. Slight variations in the amplitude of different center bursts can be observed, which are mainly attributed to airflow disturbances in the outdoor environment. Nevertheless, as illustrated in Fig. 3(b), the carrier waveforms within individual center bursts remain stable and exhibit good consistency. As shown in Fig. 3(c), the spectral SNR extracted from coherently averaged IGMs increased with the averaging time $\tau$. A log-log fit of the SNR versus $\tau$ yielded a slope of approximately 0.43. While the ideal shot-noise limit predicted an SNR scaling of $\sqrt{\tau}$, the slightly reduced slope indicated additional noise sources, such as a slow frequency drift of the CW laser and small, slow length fluctuations in the non-common optical paths of the two combs [32,33]. Figs. 3(d–f) show the measured $H_2O$ absorption spectra in the wavelength range of 3383–3386.5 nm with averaging times of 1 s, 5 s, and 30 s, respectively. As the averaging time increases, the spectral SNR improves significantly, further confirming the long-term mutual coherence of the MIR dual-comb system. Fig. 3(g) presents the broadband absorption spectrum covering the wavelength range from 3350 to 3475 nm. The blue shaded area represents the measured absorption spectrum, while the red curve indicates the fitted baseline. The baseline was obtained by fitting the measured spectrum using a combination of a third-order polynomial and trigonometric functions [34,35].

The overall dimensions of the DCS unit were 28 × 30 × 25 cm (L × W × H), corresponding to a total weight of approximately 22 kg. Together with the gas cell and photodetector, the entire setup was mounted on a 30 × 60 cm optical breadboard. The fiber-laser module and frequency-conversion module were connected through fiber flanges and collimator interfaces, allowing convenient plug-and-play operation. The power consumption was approximately 20 W, which enabled long-term continuous operation using a portable outdoor power supply. In addition, established technical standards for mobile atmospheric environmental monitoring, such as the General Technical Specification for the Mobile Laboratory of Atmospheric Environmental Monitoring (GB/T 37940-2019), provide guidance on the design and operational criteria relevant to road-safe operation, environmental tolerance (temperature and humidity), and power-supply requirements for the vehicle-mounted dual-comb system [36].

### Field-deployment measurement procedure

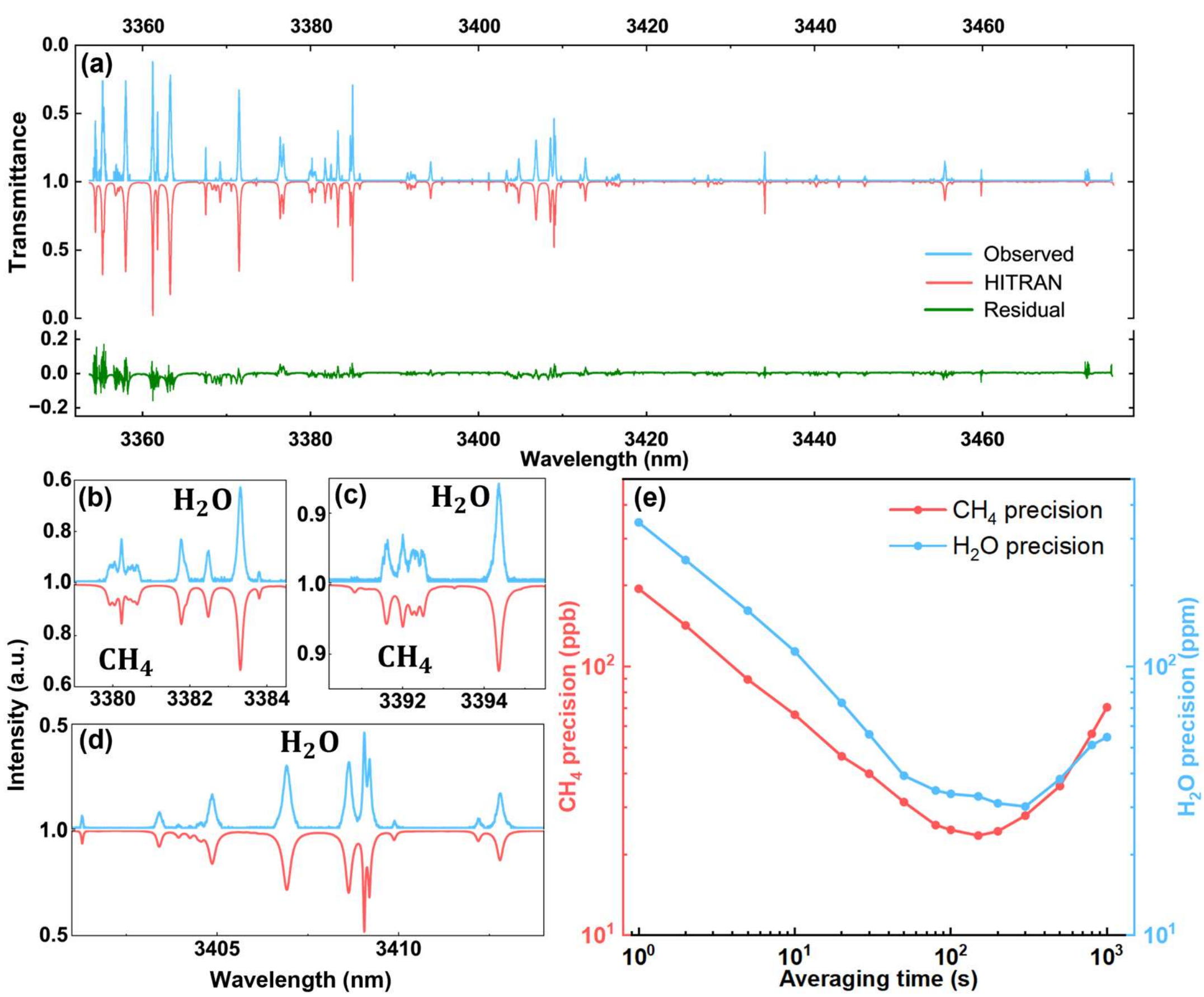

**Fig. 4**. Quantitative retrieval of $CH_4$ and $H_2O$ concentrations using the MIR dual-comb spectrometer. (a) Retrieved transmittance spectrum with a 10-s averaging time (blue), compared with the simulated spectrum based on the HITRAN database (red). The residuals of the spectral fitting are shown in green. (b–d) Expanded views of selected absorption features in different wavelength regions. The measured absorption intensities agree well with the HITRAN simulations. (e) Allan deviation analysis of the retrieved $CH_4$ and $H_2O$ mole fractions. The initial slopes ($-0.47$) indicate near white-noise-limited performance.

To evaluate the performance of the dual-comb spectrometer under field conditions, the system was operated continuously for 2000 s in an open outdoor environment. Fig. 4(a) shows the retrieved normalized transmittance spectrum with an averaging time of 10 s, compared with the simulated spectrum based on the HITRAN database. Under ambient atmospheric conditions with a $CH_4$ concentration of approximately 1.9 ppm and an optical path length of 25 m, the $CH_4$ absorption features correspond to transmittance reductions of approximately 6%–8%. The green curve at the bottom represents the residual between the measured spectrum and the HITRAN-based model. The standard deviation of the residual over the entire spectral range is 1.48%. As the spectral intensity decreases toward the edges of the comb spectrum, the SNR correspondingly degrades, leading to larger residuals in these regions. Within the central spectral range of 3370–3470 nm, the standard deviation of the residual is reduced to 0.82%. Figs. 4(b–d) present expanded views of absorption features in three representative spectral regions. The measured transmittance spectra show excellent agreement with the simulated spectra calculated from the HITRAN database under the corresponding concentration conditions. The Allan deviation of the retrieved gas concentrations was calculated to quantify the temporal stability of the system. As shown in Fig. 4(e), the initial slopes of approximately $-0.47$ for both $CH_4$ and $H_2O$ are close to the ideal slope of $-0.5$ expected for white-noise-limited performance, and are broadly consistent with the SNR scaling behavior observed in Fig. 3(c). The Allan deviations reached minimum values of 23.5 ppb for $CH_4$ and 29.6 ppm for $H_2O$ at averaging times of 150 s and 300 s, respectively. At longer integration times, slow baseline drifts became apparent. This behavior has been widely reported in laser spectroscopy systems and is generally attributed to residual interference effects caused by unintended etalon formation within the optical layout [22,37]. In the spectral domain, such interference manifests as periodic intensity modulations that distort the absorption features, particularly the sharper $H_2O$ lines, and ultimately contribute to the increase in Allan deviations at longer averaging times. These effects can be mitigated by improving thermal stability, reducing the number of parallel optical surfaces, and minimizing angular misalignment in transmissive optical components.

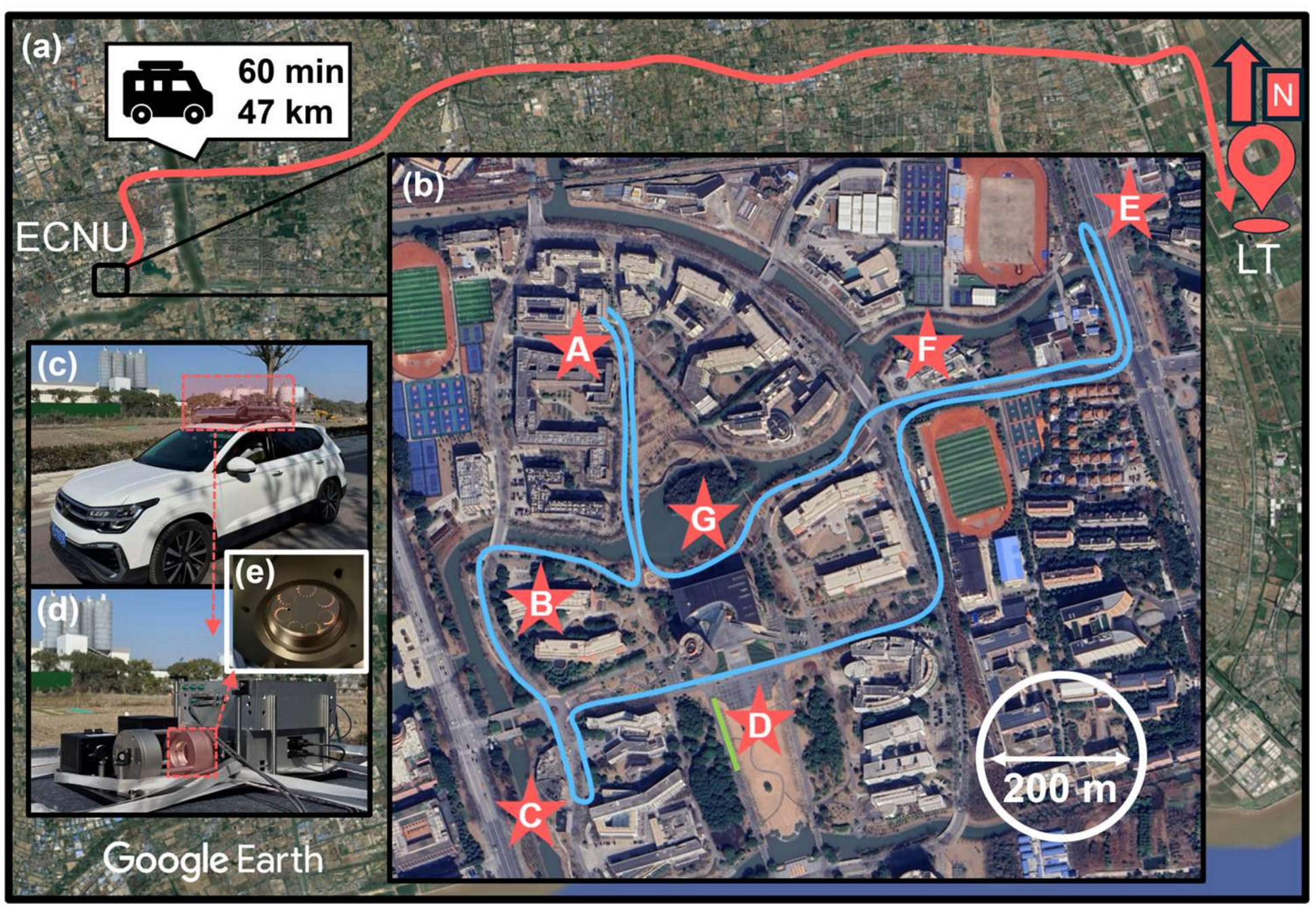

**Fig. 5**. Overview of the vehicle-mounted field-deployment experiments. (a) Map of ECNU and the surrounding urban area of Shanghai. The red trace indicates the 47 km driving route from ECNU to Laogang Town (LT). (b) Route map of the on-campus field tests. Locations A–G correspond to the Optics Building, Teaching Building, Cafeteria I, Library, Waste Compression Station, Cafeteria II, and Lakeside, respectively. A controlled natural-gas leakage patrol measurement was conducted near point D (green segment). (c) Photograph illustrating the vehicle-based field-deployment during outdoor measurements. (d) Vehicle-mounted configuration showing the DCS and open gas cell installed on the sport utility

vehicle (SUV) roof, fully exposed to ambient air. (e) Beam spots observed on one side of the Herriott gas cell.

The vehicle-mounted field experiments were conducted in two scenarios: (1) background monitoring of atmospheric $CH_4$ and $H_2O$ concentrations; and (2) localization and quantitative detection of artificially released natural-gas leakage sources. First, background patrol measurements were performed within the ECNU campus over a total distance of approximately 4 km with the vehicle speed maintained at approximately 20 km/h, as shown in Fig. 5(b). The experiments were conducted under partly cloudy conditions with occasional light gusts, and the ambient temperature ranged from 28–33 °C. During the patrol, the vehicle stopped briefly at seven designated locations, and fixed-point measurements were performed for 10 s at each site to improve the SNR. Subsequently, a long-distance mobile test was conducted along a 47 km route from ECNU to LT in the Pudong District of Shanghai on the same day, as shown in Fig. 5(a). The average driving speed across urban streets and expressways was approximately 40 km/h with instantaneous speeds occasionally reaching 100 km/h. Figs. 5(c) and 5(d) show photographs of the vehicle-mounted dual-comb system during urban patrol measurements. The optical breadboard housing the dual-comb spectrometer was secured to the SUV roof rack using elastic straps, with multilayer cushioning materials placed underneath to mitigate mechanical vibrations. An open gas cell was mounted on the front of the vehicle to enable rapid air exchange with the surrounding environment for real-time concentration measurements. Fig. 5(e) shows the visible beam generated by the nonlinear effects in the waveguide. After multiple reflections inside the Herriott gas cell, the beam formed a characteristic multi-spot pattern.

**Table 1.** Measured $CH_4$ and $H_2O$ concentrations at seven locations (A–G) within the ECNU campus.

| Mole Fraction | A | B | C | D | E | F | G |
|---|---|---|---|---|---|---|---|
| **$CH_4$ (ppm)** | 1.98 | 1.69 | 1.78 | 1.74 | 1.65 | 1.74 | 1.87 |
| **$H_2O$ (ppm)** | 27060 | 28860 | 25430 | 25880 | 26710 | 26080 | 27090 |

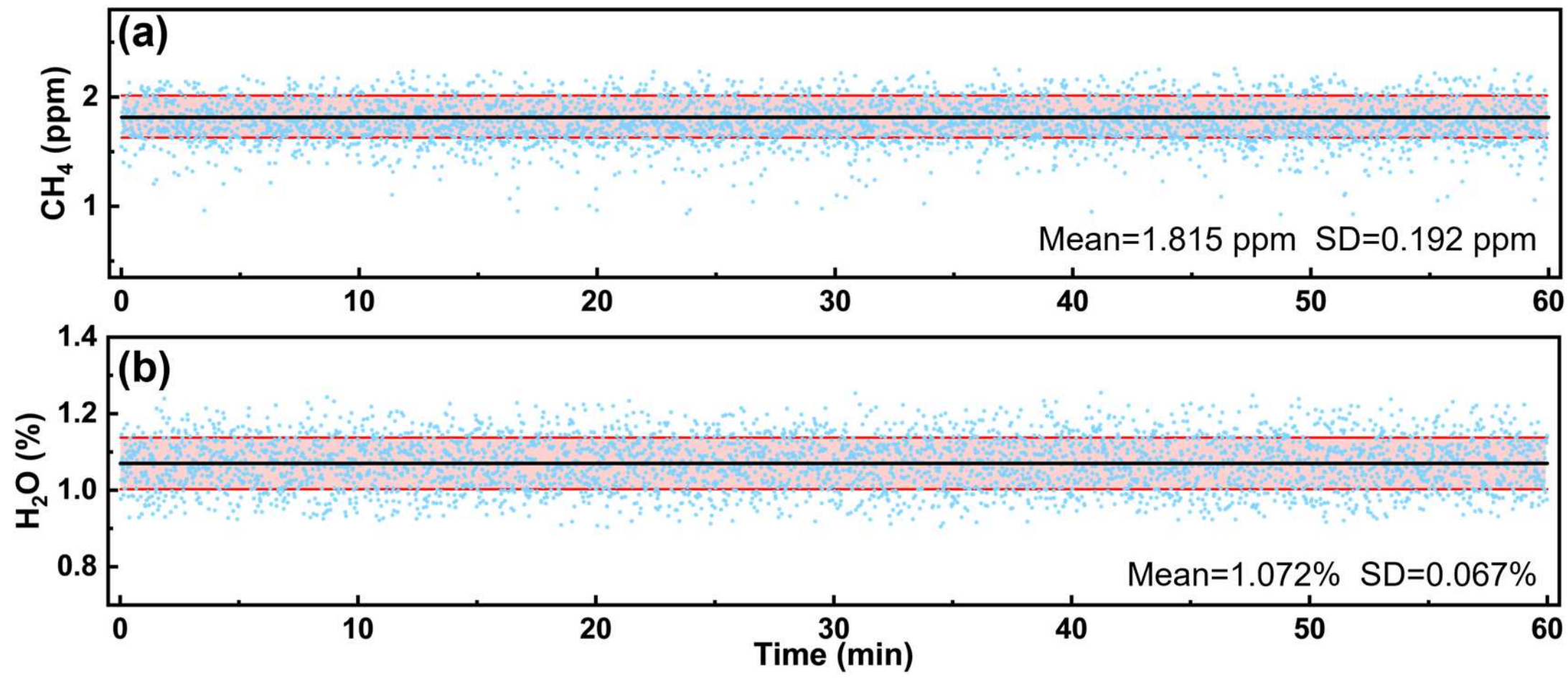

**Fig. 6.** Background concentrations of (a) $CH_4$ and (b) $H_2O$ measured during a 47 km urban patrol over 60 min. The black solid lines indicate the mean values, and the shaded regions represent ± one standard deviation (SD).

Table 1 lists the $CH_4$ and $H_2O$ mole fractions measured at seven locations (A–G) across the ECNU campus, as marked in Fig. 5(b). Each data point was retrieved from a 10 s average of the dual-comb IGMs. The gas concentrations were obtained by fitting the measured transmittance spectra to HITRAN-based reference spectra using the least-squares retrieval method [38]. The measured $CH_4$ concentrations ranged from 1.65–1.98 ppm. This range was larger than the Allan-limited precision of 66 ppb at 10 s averaging, indicating that the fluctuations were influenced not only by instrumental noise but also by small but real spatial variations in ambient $CH_4$ abundance along the campus environment. In comparison, the measured $H_2O$ concentrations spanned from 25,430 to 28,860 ppm. This spatial variation exceeded the corresponding Allan-limited precision of 114 ppm by one order of magnitude, confirming that the system clearly resolved the genuine spatial changes in atmospheric humidity. Figs. 6(a) and 6(b) show the atmospheric background $CH_4$ and $H_2O$ concentrations, respectively, measured during the 47 km urban patrol over 60 min. The concentrations were retrieved at a temporal resolution of 1 s. The mean $CH_4$ and $H_2O$ concentrations were 1.815 ppm and 1.072%, respectively, with standard deviations of 0.192 ppm and 0.067%, respectively. These results demonstrated that the background levels of both species remained stable throughout the long-distance scan, establishing a reliable baseline for identifying localized concentration enhancements.

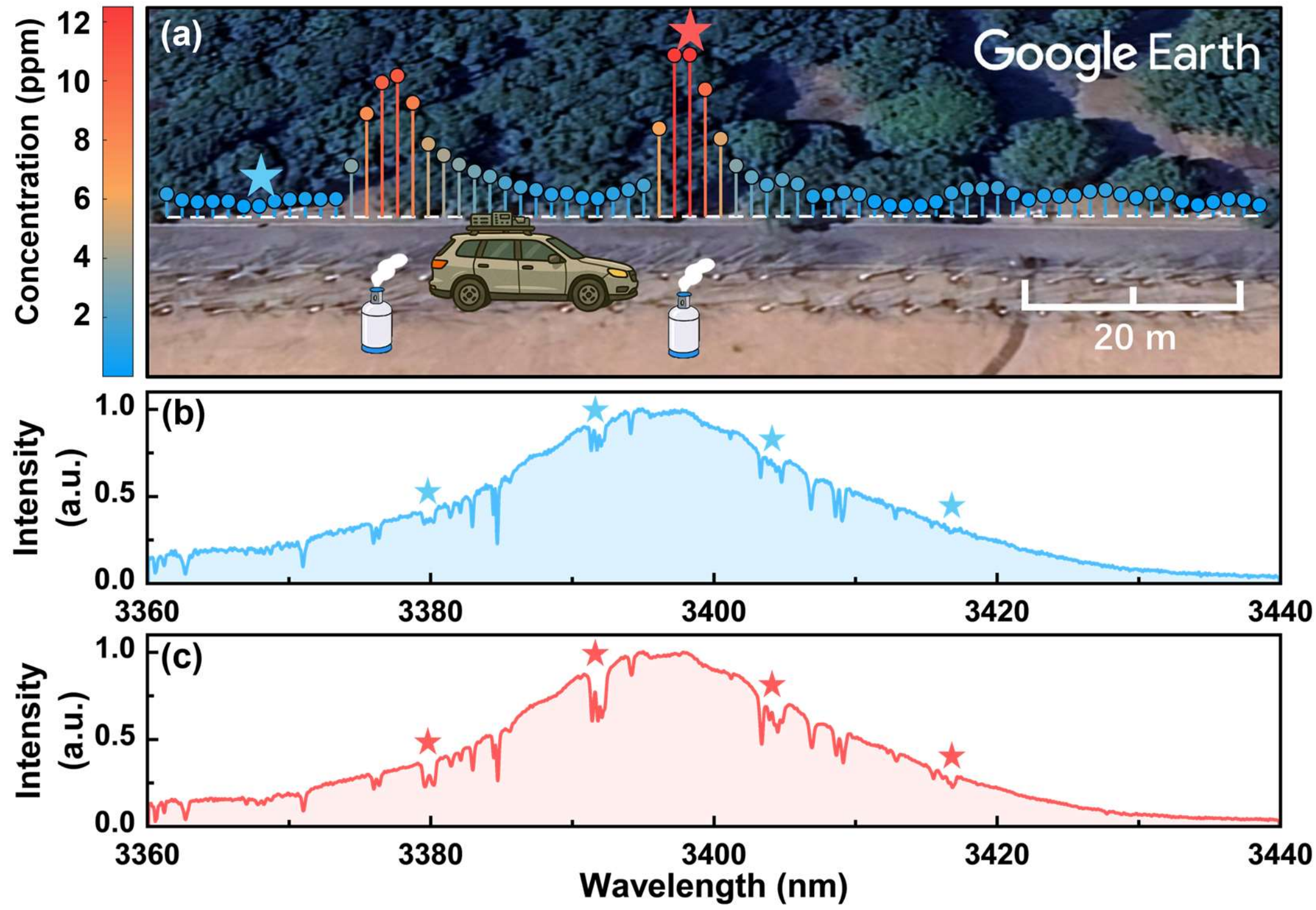

**Fig. 7**. Controlled natural-gas leakage detection experiment showing spatial $CH_4$ distributions. (a) Spatial mapping of the $CH_4$ concentration along a roadside during vehicle-based patrol measurements of two natural-gas leakage sources. The map shows 72 evenly spaced sampling points, corresponding to a 0.25 s temporal resolution at a vehicle speed of ~20 km/h over a ~100 m driving distance. (b) Representative absorption spectrum measured at a background location with a low $CH_4$ concentration, corresponding to the blue star marked in panel (a). (c) Representative absorption spectrum measured near a leakage source with a high $CH_4$ concentration, corresponding to the red star marked in panel (a). The stars in panels (b) and (c) highlight four prominent $CH_4$ absorption features.

In outdoor application scenarios, such as energy facilities and industrial sites, gas leakages typically occur as spatially confined plumes with pronounced concentration gradients. These conditions place increased demands on the spatial resolution and dynamic response of measurement techniques. After validating the system performance in ambient atmospheric background measurements, further experiments were conducted to assess its capability for flexible detection of localized gas leakages. The experiment was conducted on a quiet campus road, as indicated by the green segment near location D in Fig. 5(b). Two natural-gas cylinders (with ~95% $CH_4$ content) were deliberately deployed outdoors as artificial leakage sources, separated by approximately 30 m, and were patrolled by the DCS system, as shown in Fig. 7(a). The controlled release rates were set to 1.0 and 1.2 g/min, respectively, with each nozzle positioned approximately 0.5 m above the ground. The cylinders were located approximately 3 m perpendicular to the driving path of the vehicle. During the measurements, the ambient wind was nearly calm, allowing the spatial diffusion of $CH_4$ to be approximated as quasi-isotropic, which increased the likelihood that the expanding plume intersected the open Herriott gas cell optical path. The DCS-equipped vehicle was then driven along the road at a speed of approximately 20 km/h, covering a distance of approximately 100 m, while continuously recording the dual-comb spectra throughout the process.

Fig. 7 summarizes the results of the controlled natural-gas leakage detection experiment, highlighting the spatial concentration mapping obtained using the vehicle-mounted MIR DCS system. As shown in Fig. 7(a), the reconstructed spatial concentration map revealed two distinct $CH_4$ plumes originating from the two release points. The plume intensity gradually decreased with the distance from each cylinder, forming elongated spatial distributions influenced by the ambient air motion. Despite the relatively small release rate, the DCS system captured concentration enhancements of several ppm above the background levels within a few meters of the leakage sources. The representative absorption spectra extracted at the two spots in Fig. 7(a) are shown in Figs. 7(b) and 7(c). Compared with background spectra, the $CH_4$ absorption depth increased significantly at elevated-concentration locations, while the $H_2O$ absorption features remained nearly unchanged, indicating that the observed spectral variations were primarily attributable to $CH_4$. Overall, these results confirmed that the developed DCS system enabled reliable localization and quantification of $CH_4$ plumes from controlled leakage sources under mobile measurement conditions.

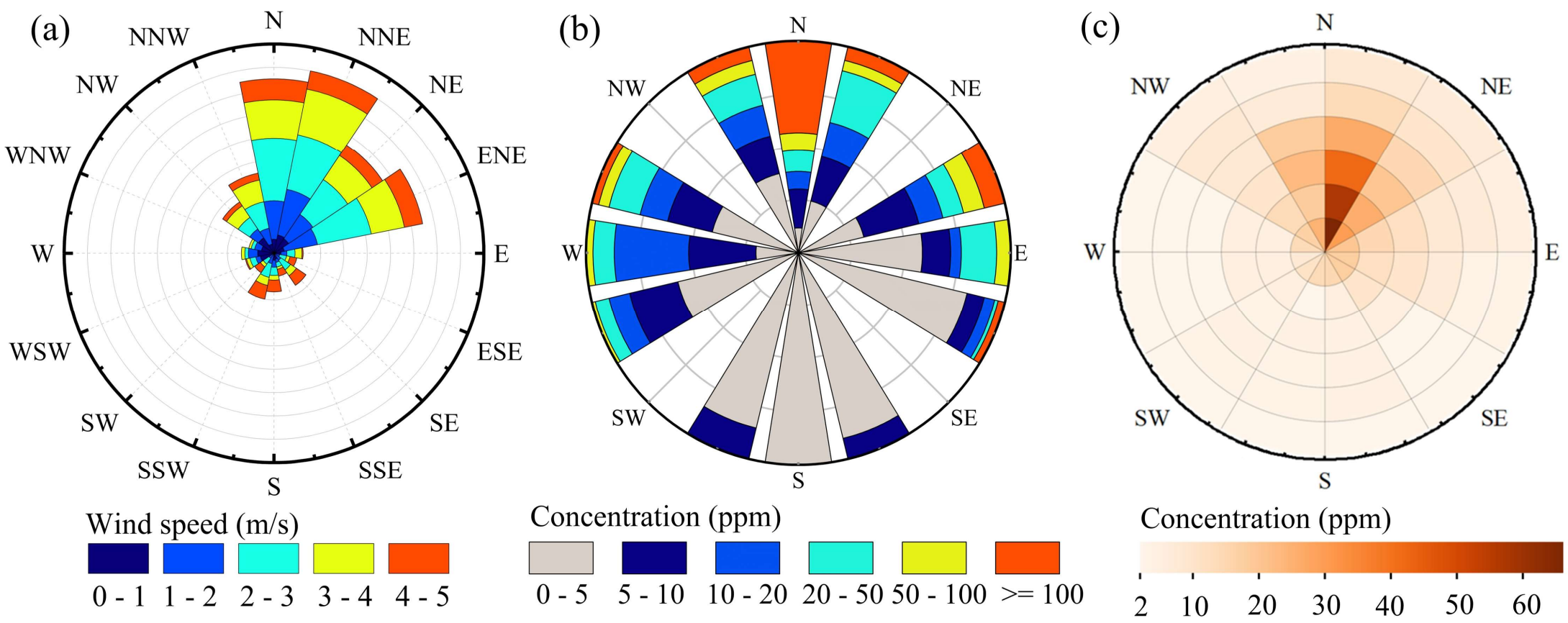

**Fig. 8.** (a) Wind-rose diagram showing the wind direction and speed distribution. (b) $CH_4$ concentration rose diagram showing the probability distribution across six concentration intervals. (c) Two-dimensional spatial concentration field map with a 1 m radial resolution.

To further investigate the spatial dispersion of $CH_4$ plumes and their dependence on wind conditions, a dedicated measurement campaign was conducted around an artificial leakage source. The spatial mapping results provided further insight into the capability of the system for quantitative plume characterization and its relationship with ambient meteorological conditions. Fig. 8 shows the spatial distribution of the $CH_4$ concentration around an artificial leakage source and the corresponding wind conditions. These experiments were conducted separately from the drive-by measurements described previously, under distinct wind conditions. The vehicle-mounted DCS system first performed measurements at twelve directions around the leakage source at a radial distance of approximately 1 m. As shown in Fig. 8(b), a concentration rose diagram was constructed by dividing each angular sector into six discrete concentration ranges. Each angular direction was measured for 60 s and the resulting time-resolved concentration data were used to derive probability distributions across discrete concentration intervals, with the radial length of each colored segment representing the corresponding probability. The pronounced anisotropy observed in the northeast direction indicated preferential plume dispersion driven by the prevailing wind, as shown in Fig. 8(a). Subsequently, measurements were carried out at 72 locations distributed at different angles and radial distances around the leakage source, with a 10-s acquisition time at each position. Fig. 8(c) presents a two-dimensional spatial concentration field map reconstructed with a 1 m radial resolution, which clearly visualizes the downwind diffusion pattern consistent with the rose diagrams in Figs. 8(a) and (b). This technique could support future studies of atmospheric transport processes in the near-surface boundary layer.

## Conclusion

In summary, this study reported the first demonstration of a vehicle-mounted DCS platform for mobile patrol measurements under field-deployment conditions. By employing an all-PM figure-9 oscillator design together with an optical-optical modulated architecture, the system achieved reduced complexity while preserving high stability and sensitivity. Atmospheric $CH_4$ and $H_2O$ concentrations were measured along a continuous driving path. In addition, controlled natural-gas leakage detection experiments were conducted, during which both leakage sources were successfully localized, and elevated $CH_4$ concentrations were continuously measured along the driving path. Furthermore, two-dimensional concentration field mapping was performed around a leakage source, revealing a strong correlation between the reconstructed $CH_4$ plume structure and the wind-rose distribution. In the future, extending the spectral coverage will enable the simultaneous detection of a broader range of gas species. Moreover, integrating the DCS system into unmanned aerial vehicle platforms can further expand its applicability and provide more flexible and wide-area solutions for practical monitoring scenarios [39–43].

### List of abbreviations

MIR: mid-infrared; DCS: dual-comb spectroscopy; VOC: volatile organic compound; PM: polarization-maintaining; ECNU: East China Normal University; CW: continuous-wave; Yb: ytterbium; OPG: optical parametric generation; FFT: fast Fourier transform; BPD: balanced photodetector; DAQ: data acquisition; IGM: interferogram; SNR: signal-to-noise ratio; LT: Laogang Town; SUV: sport utility vehicle; SD: standard deviation.

**Acknowledgements:** Not applicable.

**Authors' contributions**: C. Gu, D. Luo, and W. Li conceived the study. X. Jing, K. Wei, and C. Gu designed and built the experimental setup, and developed the software. X. Qin, J. Li, X. Yang, Z. Huang, J. Zhang, C. Sun, and C. Liu performed the experiments. X. Jing, C. Gu, and D. Luo analyzed the data and prepared visualizations. X. Jing, C. Gu, D. Luo, and W. Li drafted the manuscript. All the authors have reviewed the manuscript.

**Funding:** This study was supported in part by the National Natural Science Foundation of China (grant numbers 12134004, 62425503, 12574436, 12274141, and 12521003).

**Availability of data and materials:** The data in this paper are available from the corresponding author upon reasonable request. The code employed in this study is available from the corresponding author.

### Declarations

### Ethics approval and consent to participate

There is no ethics for this paper.

### Consent for publication

All authors agree to publish this article.

### Competing interests

The authors declare that they have no competing interests.